\documentclass[runningheads]{svmult}

\usepackage{physprbb}
\usepackage{psfig}

\begin{document}
\title*{Granular Gases - the early stage}
\toctitle{Granular Gases - the early stage}
\titlerunning{Granular Gases - the early stage}

\author{Nikolai V. Brilliantov\inst{1,2}
\and Thorsten P\"oschel\inst{3}}

\authorrunning{Nikolai V. Brilliantov and Thorsten P\"oschel}

\institute{MPIKG, Am M\"uhlenberg, 14424, Potsdam, Germany
\and Moscow State University, Physics Department,  Moscow 119899, Russia
\and Humboldt-University Berlin - Charit\'e, Institute for Biochemistry,
Hessische~Str.\,3-4, D-10115 Berlin, Germany. email: thorsten@physik.hu-berlin.de\\
http://summa.physik.hu-berlin.de/$\sim$thorsten/}

\maketitle
\vspace{-5.5cm}
\noindent {\small In: D. Reguera, L. L. Bonilla, M. Rub\'{\i} (Eds.): {\em Coherent Structures in Complex Systems}, Lecture Notes in Physics, Springer (2001), p. 408}
\vspace{4.5cm}

\begin{abstract}
  We investigate the evolution of the velocity distribution function
  of a granular gas composed of viscoelastic particles in the
  homogeneous cooling state, i.e. before clustering occurs. The
  deviation of the velocity distribution function from the Maxwellian
  distribution is quantified by a Sonine polynomials expansion. The
  first non-vanishing Sonine coefficient $a_2(t)$, reveals a complex
  time dependence which allows to assign the granular gas the property
  of an age. We discuss the possibility to measure the age of a
  granular gas.
\end{abstract}

\section{Introduction}
Granular Gases as rarefied systems of granular particles in the
absence of gravity may be exemplified by a cloud of interstellar dust.
Similar as molecular gases Granular Gases may be described within the
concepts of classical Statistical Mechanics, such as temperature $T$,
velocity distribution function $f(v)$, etc. Once initialized with a
certain velocity distribution, Granular Gases cool down due to
inelastic collisions of their particles. Although these systems are
extremely simple, in principle, they reveal a variety of structure
formation and much work has been done recently to characterize the
properties of cooling Granular Gases (see~\cite{GG} with  many
references therein).

Most of these results have been obtained under the assumption that the
coefficient of restitution $\epsilon$ which characterizes the loss of
energy of two colliding particles (see below), is a material constant.
The assumption $\epsilon=\mbox{const.}$, however, does not only
contradict experiments which show that $\epsilon$ depends significantly on
the impact velocity~\cite{CollExp,Lin}, but it contradicts even some basic
mechanical laws~\cite{rospap}. The simplest physically correct
description of dissipative particle collisions is based on the
assumption of viscoelastic material deformation during
collisions~\cite{BSHP}, which is valid for particle collisions in a
certain range of impact velocity and is in good agreement with
experimental data~\cite{Lin,Kuwabara}.

We investigate the statistical properties of Granular Gases of
viscoelastic particles for which the dependence of the restitution
coefficient on the impact velocity $\epsilon=\epsilon(v_{\rm imp})$ is
known~\cite{rospap,TomThor}. Starting from a homogeneous distribution
we study the early stage of its evolution, where no spatial
structures, as clusters~\cite{Goldhirsch,McNamara} and vortexes~\cite{Ernst}, have emerged yet. This stage is called the
homogeneously cooling state. Our results show that the
properties of Granular Gases change qualitatively if one takes into
account viscoelastic material properties, i.e.
$\epsilon=\epsilon(v_{\rm imp})$~\cite{diff}, as compared with the
equivalent system, but under the oversimplified assumption
$\epsilon=\mbox{const.}$.

In the next Section we discuss briefly the impact-velocity dependence
of the normal restitution coefficient. In Sec.~\ref{sec:epsconst}
we  introduce the method to study the velocity distribution by means of 
 the Boltzmann equation with the  Sonine polynomials expansion formalism,
and discuss the results obtained for gases of particles
interacting  with a constant restitution coefficient. Our main
results, which describe the time evolution of the velocity distribution 
for granular gases of  viscoelastic particles are derived in Sec.
\ref{sec:visco}. In Sec.~\ref{sec:age} we discuss the concept of 
the age of a granular gas which is based on the time evolution
of the velocity distribution. Finally, in conclusion we summarize 
our findings.

\section{Two-particle interaction of viscoelastic spheres}

The microscopic dynamics of granular particles is governed by the
(normal) restitution coefficient $\epsilon$ which relates the normal
components of the particle velocities before and after a collision,
$\vec{v}_{\mbox{\footnotesize\em ij}}\equiv\vec{v}_{i}-\vec{v}_{j}$
and $\vec{v}_{\mbox{\footnotesize\em
    ij}}^{*}\equiv\vec{v}_i^*-\vec{v}_j^*$ by
$\left|\vec{v}_{\mbox{\footnotesize\em ij}}^{*} \vec{e}\right| =
\epsilon \left|\vec{v}_{\mbox{\footnotesize\em ij}} \vec{e}\right|$.
The unit vector $\vec{e}=\vec{r}_{\mbox{\footnotesize\em
    ij}}/\left|\vec{r}_{\mbox{\footnotesize\em ij}}\right|$ gives the direction
of the inter-center vector $\vec{r}_{\mbox{\footnotesize\em
    ij}}=\vec{r}_i-\vec{r}_j$ at the instant of the collision. From
the conservation of momentum one finds the change of velocity for the
colliding particles:
\begin{equation}
\vec{v}_{i}^{*} = \vec{v}_{i} -\frac12 \,(1+\epsilon)
\left(\vec{v}_{\mbox{\footnotesize\em ij}} \cdot  \vec{e}\,\right) 
\vec{e}\,,\,\,\,\,\,\,\,\,\,\,\,\,\,\,\,
\vec{v}_{j}^{*} = \vec{v}_{j} +\frac12 \,(1+\epsilon)
\left(\vec{v}_{\mbox{\footnotesize\em ij}} \cdot  \vec{e}\,\right) 
\vec{e}\, .
\label{eq:restdef}
\end{equation}
For elastic collisions one has $\epsilon =1$ and for inelastic
collisions $\epsilon $ decreases with increasing degree of
inelasticity.

In literature it is frequently assumed that the restitution
coefficient is a material constant, $\epsilon=\mbox{const.}$
Experiments, e.g.~\cite{CollExp,Lin}, as well as theoretical
investigations~\cite{BSHP} show, however, that this assumption is not
consistent with the nature of inelastic collisions, it does not agree
even with a dimension analysis~\cite{rospap}. The impact velocity
dependence of the restitution coefficient $\epsilon\left(v_{\rm imp}
\right)= \epsilon\left(\left|\vec{e}\, \vec{v}_{\mbox{\footnotesize\em
        ij}}\right|\right)$ has been obtained by generalizing Hertz's
contact problem to viscoelastic spheres~\cite{BSHP}. From the
generalized Hertz equation~\cite{Hertz} one obtains the
velocity-dependent restitution coefficient for viscoelastic
spheres~\cite{TomThor}
\begin{equation}
\label{epsC1C2}
\epsilon=1- C_1 A \alpha^{2/5}\left|\vec{e}\,
\vec{v}_{\mbox{\footnotesize\em ij}}\right|^{1/5}
+C_2 A^2  \alpha^{4/5}\left|\vec{e}\,
\vec{v}_{\mbox{\footnotesize\em ij}}\right|^{2/5}  \mp  \cdots
\label{epsilon}
\end{equation}
with
\begin{equation}
\alpha= \left( \frac32 \right)^{3/2}
\frac{ Y\sqrt{R^{\,\mbox{\footnotesize eff}}}}{
    m^{\mbox{\footnotesize eff}}\left( 1-\nu ^2\right) }\,,
\end{equation}
where $Y$ is the Young modulus, $\nu$ is the Poisson ratio, and $A$
depends on dissipative parameters of the particle material (for
details see~\cite{BSHP}). The effective mass and radius are defined as
\begin{equation}
R^{\,\mbox{\footnotesize eff}}=R_1R_2/(R_1+R_2)~~~~~~~~~~
m^{\mbox{\footnotesize eff}}=m_1m_2/(m_1+m_2)
\end{equation}
with $R_{1/2}$ and $m_{1/2}$ being the radii and masses of the
colliding particles. The constants are given by~\cite{rospap,TomThor}
\begin{equation}
C_1=\frac{ \Gamma(3/5)\sqrt{\pi}}{2^{1/5}5^{2/5} \Gamma(21/10)}
\approx 1.1534 ~~~~~~~~~~
C_2=\frac35C_1^2\approx 0.7982 \,.
\label{C2}
\end{equation}
Equation (\ref{epsilon}) refers to the case of pure viscoelastic
interaction, i.e. when the relative velocity $\left|\vec{v}_{ij}\vec{e}\right|$
is not too large (to avoid plastic deformation of the particles) and
is not too small (to allow to neglect surface effects such as
roughness, adhesion and van der Waals interactions). The dependence of
$\epsilon=\epsilon\left(\left|\vec{e}\vec{v}_{ij}\right|\right)$
(without the material dependence) was already mentioned in
\cite{Kuwabara} where heuristic arguments have been applied.

In what follows we consider a granular gas of identical viscoelastic
spheres of unit mass.

\section{Kinetics of Granular Gases: The case $\epsilon=\mbox{const.}$}
\label{sec:epsconst}

The evolution of the velocity distribution function is generally
described by the Boltzmann-Enskog equation, which for the force-free
case reads~\cite{Esipov,NoijeErnst:97,resibua}:
\begin{eqnarray}
\label{collint}
&&\frac{\partial}{\partial t}f\left(\vec{v},t\right) =
g_2(\sigma)\sigma^2 \int d \vec{v}_2 \int d\vec{e}
\Theta\left(-\vec{v}_{12} \cdot \vec{e}\right) \left|\vec{v}_{12} \cdot \vec{e}\right| \nonumber \\
&& ~~~~\times \, \left\{\chi f(\vec{v}_1^{**},t)f(\vec{v}_2^{**},t)-
f(\vec{v}_1,t)f(\vec{v}_2,t) \right\} \equiv g_2(\sigma) I(f, f)\, ,
\end{eqnarray}
where $\sigma=2R $ is the diameter of particles,
$g_2(\sigma)=(2-\eta)/2(1-\eta)^3$ ($\eta=\frac16\, \pi n \sigma^3$ is
packing fraction) denotes the contact value of the two-particle
correlation function~\cite{CarnahanStarling}, which accounts for the
increasing collision frequency due to the excluded volume effects;
$\Theta(x)$ is the Heaviside step-function. The velocities $\vec{v}_1^{**}$ and $\vec{v}_2^{**}$ refer to the precollisional
velocities of the so-called inverse collision, which results with
$\vec{v}_1$ and $\vec{v}_2$ as the after-collisional velocities.
Finally the factor $\chi$ in the gain term appears respectively from
the Jacobian of the transformation $d\vec{v}_1^{**}d\vec{v}_2^{**} \to
d\vec{v}_1 d\vec{v}_2$ and from the relation between the lengths of
the collisional cylinders $\epsilon \left|\vec{v}_{12}^{**} \cdot
  \vec{e}\right| dt=\left|\vec{v}_{12} \cdot \vec{e}\right|dt$.
\cite{veldistr}. For the constant restitution coefficient $\epsilon =
{\rm const.}$ this  reads $\chi=\epsilon^{-2}$
\cite{NoijeErnst:97,GoldshteinShapiro95}, i.e., it is independent on
the impact velocity and, therefore, independent on time.

With  the scaling Ansatz for the distribution function
\begin{equation}
\label{veldis}
f(\vec{v}, t)=\frac{n}{v_0^3(t)} \tilde{f} \left( \frac{v}{v_0(t)}\right)
=\frac{n}{v_0^3(t)} \tilde{f} (\vec{c})   \,,
\end{equation}
where $n$ is the number density of the granular gas and $v_0(t)$ is
the thermal velocity defined by
\begin{equation}
\label{deftemp1}
\frac{3}{2} n T(t)= \int d \vec{v} \frac{v^2}{2} f(\vec{v},t)\,,
=\frac32 n v_0^2(t)
\end{equation}
the Boltzmann equation may be reduced to two independent equations:
one for the (time-independent) scaling function $\tilde{f} (\vec{c})$, and the 
other one for the time-dependence of the thermal velocity (i.e. for the 
temperature). Solving the equation for  the temperature one obtains
$T(t)=T_0/\left(1+t/\tau\right)^2$~\cite{Esipov}. 

The solution of the other equation for $\tilde{f} (\vec{c})$, may be found 
in terms of the Sonine polynomial
expansion~\cite{NoijeErnst:97,GoldshteinShapiro95,EpsConst}. For the 
case  of $\epsilon=\mbox{const.}$ it reads
\begin{equation}
\label{Soninexp}
\tilde{f}(\vec{c})=\phi(c) \left\{1 + \sum_{p=1}^{\infty}
a_p  S_p(c^2) \right\}\,,
\end{equation}
where $\phi(c) \equiv \pi^{-d/2} \exp(-c^2)$
 is the Maxwellian distribution for the rescaled velocity, and the first 
few Sonine polynomials are
\begin{equation}
\label{Soninfewfirst}
S_0(x)=1,~~~~
S_1(x)=-x^2 +\frac32,~~~~~
S_2(x)=\frac{x^2}{2}-\frac{5x}{2}+\frac{15}{8}\,.
\end{equation}
The leading (zero-order) term in Eq. (\ref{Soninexp}) is the
Maxwellian distribution, while the next-order terms, characterized by
the coefficients $a_i$ describe the deviation of the distribution from
the Maxwellian. For the case of $\epsilon={\rm const.}$ the velocity
distribution function $\tilde{f}(\vec{c})$  is time-independent, therefore, the coefficients
of the Sonine polynomials expansion are constants.

If the inelasticity is small, one can restrict to the first
non-vanishing term beyond the Maxwellian which has the coefficient $a_2$ 
($a_1=0$, according to the definition of temperature~\cite{NoijeErnst:97,GoldshteinShapiro95,EpsConst}). For
$\epsilon=\mbox{const.}$ this coefficient reads~\cite{NoijeErnst}
\begin{equation}
  \label{eq:a2Noije}
a_2=\frac{16(1-\epsilon)(1-2\epsilon^2)}{9+24d
+8\epsilon d +41\epsilon+30(1-\epsilon)\epsilon^2 }\,.
\end{equation}
A more accurate expression for $a_2$ may be found in~\cite{EpsConst}.

\section{Kinetics  of Granular Gases: Viscoelastic particles}
\label{sec:visco}

For viscoelastic particles the restitution coefficient $\epsilon$
depends on the impact velocity due to Eq.~(\ref{epsC1C2}). Hence, the
factor $\chi$ in the Boltzmann equation (\ref{collint}) is not anymore
constant as for $\epsilon=\mbox{const.}$ but it reads
\begin{equation}
\label{CHI}
\chi = 1 + \frac{11}{5}C_1 A \alpha^{2/5} \left|\vec{v}_{12} \cdot
\vec{e}\right|^{1/5} +\frac{66}{25}C_1^2 A^2 \alpha^{4/5}
\left|\vec{v}_{12} \cdot \vec{e}\right|^{2/5} +\cdots
\end{equation}
Using again the scaling Ansatz (\ref{veldis}) the rhs of the Boltzmann equation does not factorize
into two parts, one depending only on the scaling function $\tilde{f}(\vec{c})$, and another one, depending only on $v_0$.
Consequently, one can not obtain a set of decoupled equations for $\tilde{f}(\vec{c})$ and for the temperature. 

Nevertheless, it is worth to substitute the generalized scaling Ansatz
\begin{equation}
\label{genscal}
f(\vec{v}, t)=\frac{n}{v_0^3(t)}\tilde{f}(\vec{c}, t)
\end{equation}
into the kinetic equation (\ref{collint}). After some 
algebra Eq.(\ref{collint}) may be recast into the form
\begin{equation}
\label{geneqveldis}
\frac{\mu_2}{3}
\left(3 + c_1 \frac{\partial}{\partial c_1} \right)
\tilde{f}(\vec{c}, t) +
B^{-1} \frac{\partial}{\partial t} \tilde{f}(\vec{c}, t) =
\tilde{I}\left( \tilde{f}, \tilde{f} \right)
\end{equation}
where we define the dimensionless collisional integral
\begin{equation}
\label{dimlcolint}
\!\!\tilde{I}\!\left(\! \tilde{f}, \tilde{f} \right)\!=\!\!
\int\!\!\! d \vec{c}_2\!\! \int\!\! d\vec{e}\,
\Theta\!\left(-\vec{c}_{12} \!\cdot\! \vec{e}\right)\!
\left|\vec{c}_{12} \!\cdot\! \vec{e}\right|
 \left\{\tilde{\chi} \tilde{f}(\vec{c}_1^{**}\!,t)
\tilde{f}(\vec{c}_2^{**}\!,t)-
\tilde{f}(\vec{c}_1,t)\tilde{f}(\vec{c}_2,t) \right\}
\end{equation}
with the reduced factor $\tilde{\chi}$
\begin{equation}
\label{chiscal}
\tilde{\chi} \!=\! 1 \!+ \!
\frac{11}{5}C_1 \delta^{\prime} \left|\vec{c}_{12}
\cdot \vec{e}\right|^{1/5}
+\frac{66}{25}C_1^2 \delta^{\prime \, 2}
\left|\vec{c}_{12} \cdot \vec{e}\right|^{2/5}
+\cdots
\end{equation}
which depends now on time via a quantity
\begin{equation}
\label{deltaprime}
\delta^{\, \prime} (t) \equiv A \alpha^{2/5} \left[2T(t)\right]^{1/10}
\equiv \delta \left[2T(t)/T_0 \right]^{1/10}
\end{equation}
Here $\delta \equiv A \alpha^{2/5}[T_0]^{1/10}$ and $T_0$ is the
initial temperature. We also define $B=B(t) \equiv v_0(t) g_2(\sigma)
\sigma^2 n$, and the moments of the dimensionless collision integral
\begin{equation}
\label{mup}
\mu_p \equiv - \int d \vec{c}_1 c_1^{\,p}
\tilde{I}\left( \tilde{f}, \tilde{f} \right)\ .
\end{equation}
According to the definitions (\ref{deftemp1}), (\ref{mup}) and of $B$,
the second moment $\mu_2$ defines the evolution of the temperature:
\begin{equation}
\label{dTdt}
\frac{dT}{dt}=\frac13g_2(\sigma) \sigma^2 n v_0^3
\int d \vec{c}_1 c_1^2
\tilde{I}\left( \tilde{f}, \tilde{f} \right) =-\frac23 BT\mu_2\,.
\end{equation}

The velocity distribution function we again describe by a Sonine
polynomials expansion as introduced in Eq.~(\ref{Soninexp}). Since in
contrast to the case $\epsilon=\mbox{const.}$, the Boltzmann equation
for a gas of viscoelastic particles does not factorize into a time
dependent equation for temperature and a time-independent equation for
the velocity distribution, the Sonine coefficients $a_i$ are not
constants but depend explicitely on time, i.e. one has now
\begin{equation}
\label{Soninexp1}
\tilde{f}(\vec{c},t)=\phi(c) \left\{1 + \sum_{p=1}^{\infty}
a_p(t)  S_p(c^2) \right\}\,.
\end{equation}

Equations for $a_p (t)$ may be 
found by multiplying both sides of Eq.(\ref{geneqveldis}) with $c_1^{\,p}$
and integrating over $d \vec{c}_1$. One obtains
\begin{equation}
\label{momeq}
\frac{\mu_2}{3} p \left< c^{\,p} \right> -B^{-1}\sum_{k=1}^{\infty}
\dot{a}_k \nu_{kp} = \mu_p
\end{equation}
where integration by parts has been performed and where we define
\begin{equation}
\label{nukp}
\nu_{kp} \equiv \int \phi(c) c^{\,p} S_k(c^2) d\vec{c}\, ; \quad
\left< c^{\,p} \right>  \equiv \int c^{\,p} \tilde{f}(\vec{c}, t)  d\vec{c} \, .
\end{equation}

{F}rom (\ref{momeq}) we see already that the granular temperature and
the Sonine coefficients and, hence, the distribution function do not
evolve independently.

The calculation of $\nu_{kp}$ is straightforward; the first few of
them read $\nu_{22}=0$, $\nu_{24}=\frac{15}{4}$.  The odd moments
$\left< c^{2n+1} \right> $ are zero, while the even ones, $\left<
  c^{2n} \right> $ may be expressed in terms of $a_k$ with $0 \leq k
\leq n$. Calculations show that $\left< c^2 \right> = \frac32$,
implying $a_1=0$, according to the definition of temperature
(\ref{deftemp1}) (e.g.\cite{NoijeErnst:97}), and that $\left< c^4
\right> = \frac{15}{4}\left( 1 + a_2 \right)$.

Now we assume that the dissipation is not large so that the deviation
from the Maxwellian distribution may be sufficiently described by the
second term in the expansion (\ref{Soninexp}) only, with all
higher-order terms with $p>2$ discarded. Then (\ref{momeq}) is an
equation for the coefficient $a_2$.  Using the above results for
$\nu_{22}$, $\nu_{24}$, $\left< c^2 \right>$ and $\left< c^4 \right>$
it is easy to show that Eq.(\ref{momeq}) converts for $p=2$ into
identity, while for $p=4$ it reads:
\begin{equation}
\label{eqa2}
\dot{a}_2-\frac43\, B\mu_2 \left(1+a_2 \right)+\frac4{15} \, B\mu_4 =0\,.
\end{equation}
With the approximation $\tilde{f}= \phi (c) [ 1+a_2(t)S_2(c^2)]$ the
time-dependent coefficients $\mu_p(t)$ may be expressed in terms of
$a_2$ due to the definition (\ref{mup}). Using the properties of the
collision integral (e.g. the conservation of the total momentum at
collision) one can obtain relations for the $\mu_p(t)$ (e.g.
\cite{NoijeErnst:97}):
\begin{eqnarray}
\label{mupa2}
&&\mu_p=-\frac12 \int d\vec{c}_1\int d\vec{c}_2 \int d\vec{e}
\Theta(-\vec{c}_{12} \cdot \vec{e}) \left|\vec{c}_{12} \cdot
\vec{e}\right| \phi(c_1) \phi(c_2)
\times \nonumber \\
&&~~~~~~~~\left\{1+a_2\left[S_2(c_1^2)+S_2(c_2^2) \right] +
a_2^2\,S_2(c_1^2)S_2(c_2^2) \right\}
\Delta (c_1^{\,p}+c_2^{\,p}) \nonumber
\end{eqnarray}
where $\Delta \psi(\vec{c}_i) \equiv
\left[\psi(\vec{c}_i^*)-\psi(\vec{c}_i) \right]$ denotes the change of
some function $\psi( \vec{c}_i)$ in a direct collision.  Calculations
up to the second order in $\delta$ yield~\cite{veldistr}
\begin{equation}
\label{mu2A}
\mu_2= \sum_{k=0}^{2} \sum_{n=0}^2 {\cal A}_{kn} \delta^{\, \prime \,k } a_2^n
\end{equation}
where the coefficients ${\cal A}$ are pure numbers:
${\cal A}_{00}=0$, ${\cal A}_{01}=0$, ${\cal A}_{02}=0 $,
${\cal A}_{10}=\omega_0$, ${\cal A}_{11}=\frac{6}{25}\omega_0$,
${\cal A}_{12}=\frac{21}{2500} \omega_0 $, ${\cal A}_{20}=\omega_1$,
${\cal A}_{21}=\frac{119}{400}\omega_1$ and
${\cal A}_{22}=\frac{4641}{640000}\omega_1 $,
with
$\omega_0 \equiv 2 \sqrt{2 \pi} 2^{1/10} \Gamma \left (\frac{21}{10} \right)C_1
\approx 6.48562$
and
$\omega_1 \equiv \sqrt{2 \pi} 2^{1/5} \Gamma \left (\frac{16}{5} \right)
C_1^2\approx 9.28569$.  Similarly
\begin{equation}
\label{mu4A}
\mu_4= \sum_{k=0}^{2} \sum_{n=0}^2 {\cal B}_{kn}
\delta^{\, \prime \,k } a_2^n
\end{equation}
with ${\cal B}_{00}=0$, ${\cal B}_{01}=4\sqrt{2 \pi}$,
${\cal B}_{02}=\frac{1}{8}\sqrt{2 \pi} $,
${\cal B}_{10}= \frac{56}{10}\omega_0$,
${\cal B}_{11}=\frac{1806}{250}\omega_0$,
${\cal B}_{12}=\frac{567}{12500}\omega_0 $,
${\cal B}_{20}= \frac{77}{10}\omega_1$,
${\cal B}_{21}=\frac{149054}{13750}\omega_1$ and
${\cal B}_{22}=\frac{348424}{5500000}\omega_1 $

Thus, Eqs.(\ref{eqa2}) and (\ref{dTdt}), together with
Eqs.(\ref{mu2A}) and (\ref{mu4A}) form a closed set to find the time
evolution of the temperature and of the coefficient $a_2$.  In
contrast to the case of constant restitution coefficient where
$a_2={\rm const.}$, in a gas of viscoelastic particles the time
evolution of temperature is coupled to the time evolution of the
Sonine coefficient $a_2$. This coupling may lead to a rather peculiar
time-evolution of the system.

Introducing the reduced temperature $u(t) \equiv T(t)/T_0$ we recast
the set (\ref{eqa2}), (\ref{dTdt}) into the form
\begin{eqnarray}
\label{genseteq1}
&&\!\!\!\dot{u}\!+\!\frac{1}{\tau_0}u^{\frac85}\!\!
\left(\! \frac53\! +\!\frac25 a_2\!+\!
\frac{7}{500}a_2^2\right)\!-\! \frac{1}{\tau_0}q_1
\delta \, u^{\frac{17}{10}}
\left( \frac53 \!+\!\frac{119}{240}a_2 \!+\!
\frac{1547}{128000}a_2^2 \right)\! =\!0\\
\label{genseteq2}
&&\!\!\!\dot{a}_2-r_0u^{1/2}\mu_2 \left(1 +a_2 \right) +
\frac15 r_0u^{1/2}\mu_4=0
\end{eqnarray}
where we introduce the characteristic time
\begin{equation}
\label{tau0}
\tau_0^{-1}=\frac{16}5 q_0 \delta \cdot \tau_c(0)^{-1}=
\frac{16}5q_0 \delta \left( 4 \sqrt{\pi} g_2(\sigma)
\sigma^2 n \sqrt{T_0} \right)^{-1}
\end{equation}
and define $q_0=2^{1/5}\Gamma(21/10)C_1/8\approx 0.173318$, 
$q_1 \equiv 2^{1/10}
(\omega_1/\omega_0) \approx 1.53445$, and 
$r_0^{-1} \equiv (24 \sqrt{2 \pi}/5)q_0 \delta \tau_0$.
Equation (\ref{tau0}), as shown
below, describes the time evolution of the temperature.  To obtain the
last equations we use the expressions for $\mu_2$, $\mu_4$, $B$, and
for the coefficients ${\cal A}$. Note that the characteristic time
$\tau_0$ is $\delta^{-1} \gg 1$ times larger than the mean collision
time $\tau_c(0)$.

We will find the solution to these equations as expansions in terms of
the small dissipative parameter $\delta$ ($\delta^{\, \prime}(t) =
\delta \cdot 2^{1/10}u ^{1/10}(t)$):
\begin{equation}
\label{expudel}
u=u_0+ \delta \cdot u_1 +\delta^2 \cdot u_2 +\cdots\,, ~~~~~~
a_2=a_{20}+\delta \cdot a_{21}+\delta^2 \cdot a_{22} +\cdots
\end{equation}
Substituting Eqs. (\ref{expudel}), (\ref{mu2A}) and (\ref{mu4A}) into
Eqs. (\ref{genseteq1}), (\ref{genseteq2}), one can solve these
equations perturbatively for each order of $\delta$.

\begin{figure}[htbp]
\centerline{\psfig{figure=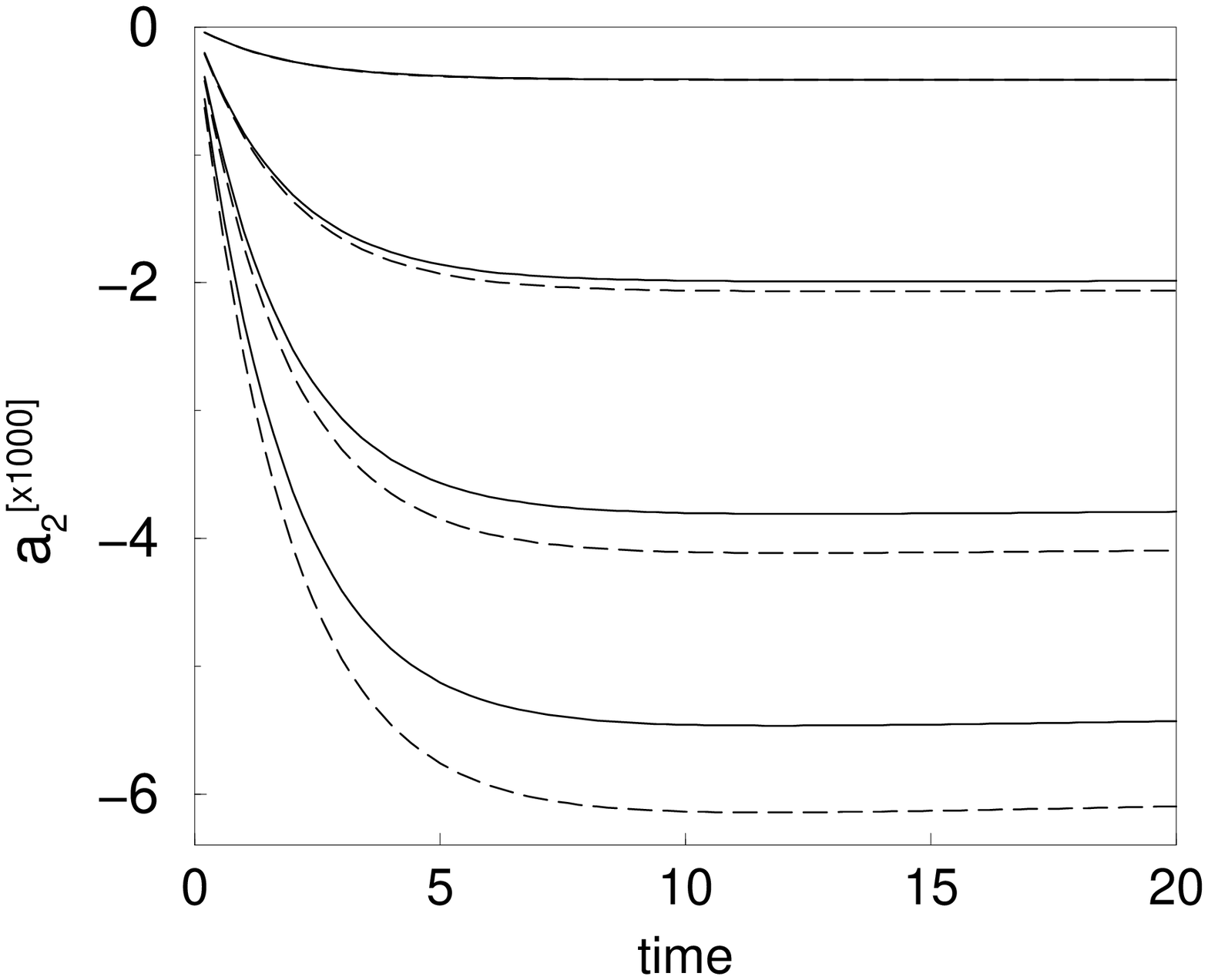,width=5.5cm}\psfig{figure=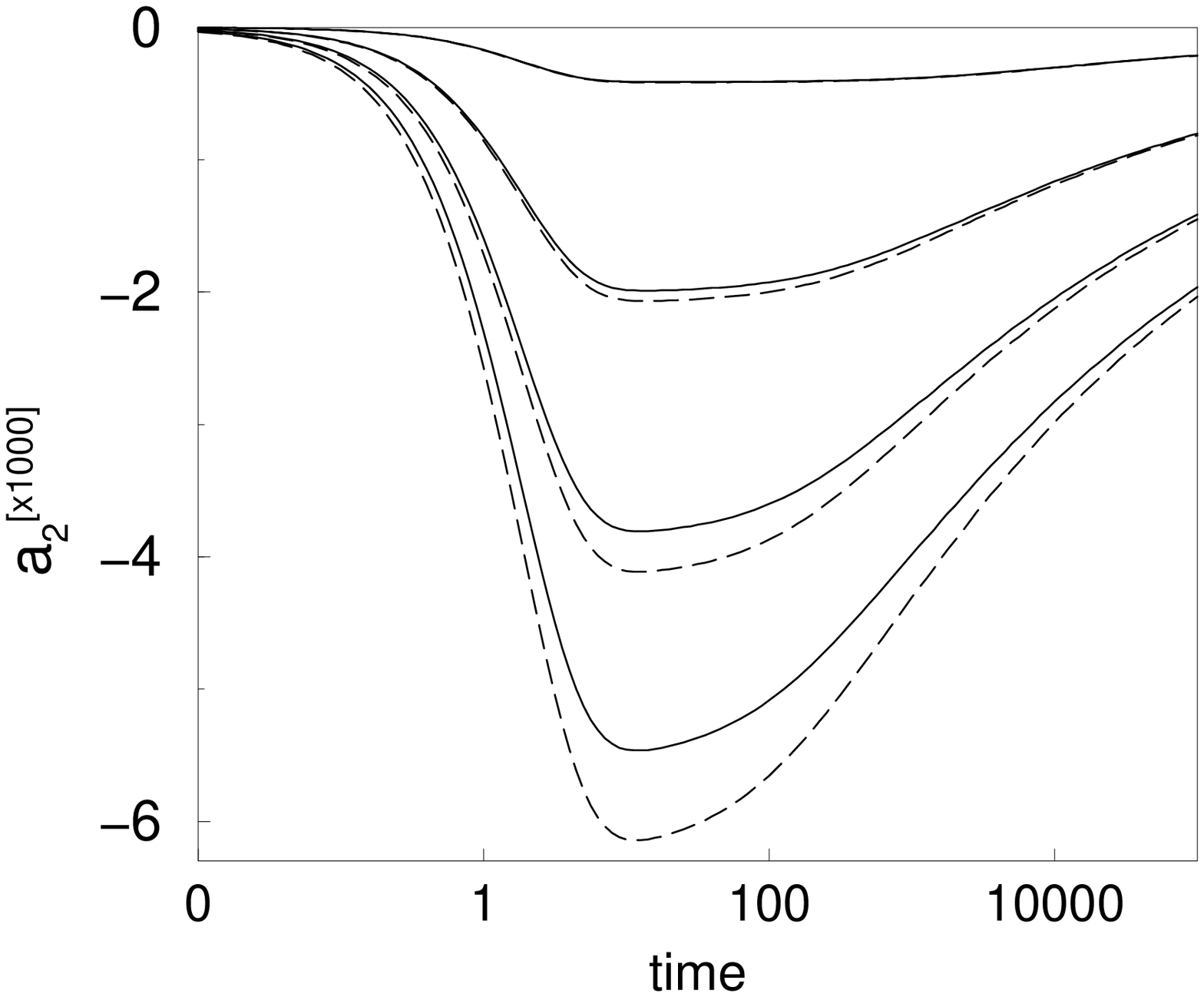,width=5.5cm}}
\centerline{\psfig{figure=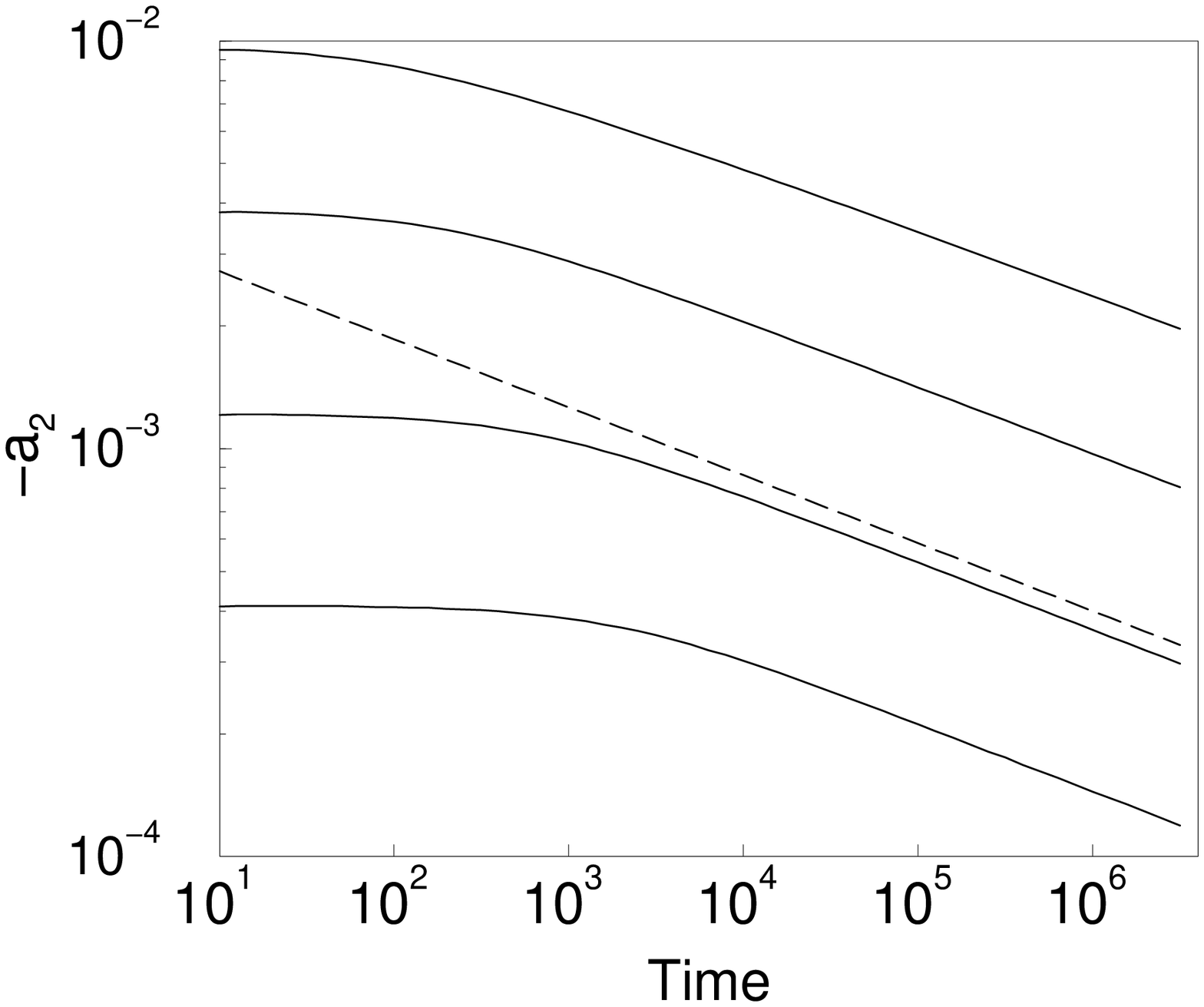,width=5.5cm}}
 \caption{Time dependence of the second coefficient of the Sonine polynomial
expansion $a_2(t) \times 100 $. Time is given in units of the mean collision
time $\tau_c(0)$. $\delta=0.1, 0.11, 0.12, \ldots, 0.20$ (bottom to top).}
\label{fig:a2small}
\end{figure}

Keeping only linear terms with respect to $\delta$, one can find the
analytical solution (see~\cite{veldistr} for details). This reads e.g.
for the coefficient $a_2(t)$:
\begin{equation}
\label{a21gensolLi}
a_2(t)=\delta\! \cdot\! a_{21}=-\frac{12}{5}w(t)^{-1}
\left\{ {\rm Li} \left[ w(t) \right]-{\rm Li} \left[ w(0) \right] \right\}
\end{equation}
where $w(t) \equiv \exp \left[ \left(q_0 \delta \right)^{-1}
  \left(1+t/\tau_0 \right)^{1/6} \right]$ and ${\rm Li}(x)$ is the
logarithmic integral.  For $t \ll \tau_0$ Eq.(\ref{a21gensolLi})
reduces to
\begin{equation}
\label{a21tsmallsol}
a_{2}(t)=
- \delta \cdot h \left( 1- e^{-8t/15\tau_c(0)} \right)
\end{equation}
where $h \equiv 2^{1/10} \left( {\cal B}_{10}-5{\cal A}_{10}\right)/16
\pi =0.415964$.  As it follows from Eq.(\ref{a21tsmallsol}), after a
transient time of the order of few collisions per particle, i.e. for
$\tau_c(0) < t \ll \tau_0$, $a_{2}(t)$ saturates at the
``steady-state'' (on the time-scale $\sim \tau_c(0)$) value
$-h\,\delta =-0.415964 \delta$. For $t \gg \tau_0$ it decays on a
``slow'' time-scale $\sim \tau_0$:
\begin{equation}
\label{a21tlargesol}
a_{2}(t) \simeq
- \delta \cdot h \left( t/\tau_0  \right)^{-1/6}
\end{equation}
and the velocity distribution tends to the Maxwellian. Linear theory
gives for the temperature for $t \gg \tau_0$~\cite{veldistr}
\begin{equation}
\label{T(t)del1}
\frac{T(t)}{T_0}= \left(1 + \frac{t}{\tau_0} \right)^{-5/3}+
\left( \frac{12}{25}h+2q_1 \right)\left(\frac{t}{\tau_0}\right)^{-11/6}
\end{equation}
with the constants $h$ and $q_1$ given above. The linear theory agrees
fairly well with numerical solution (see Fig.~\ref{fig:a2small} where the
time dependence of $a_2$ is given) for small $\delta$.

\begin{figure}[htbp]
  \centerline{\psfig{figure=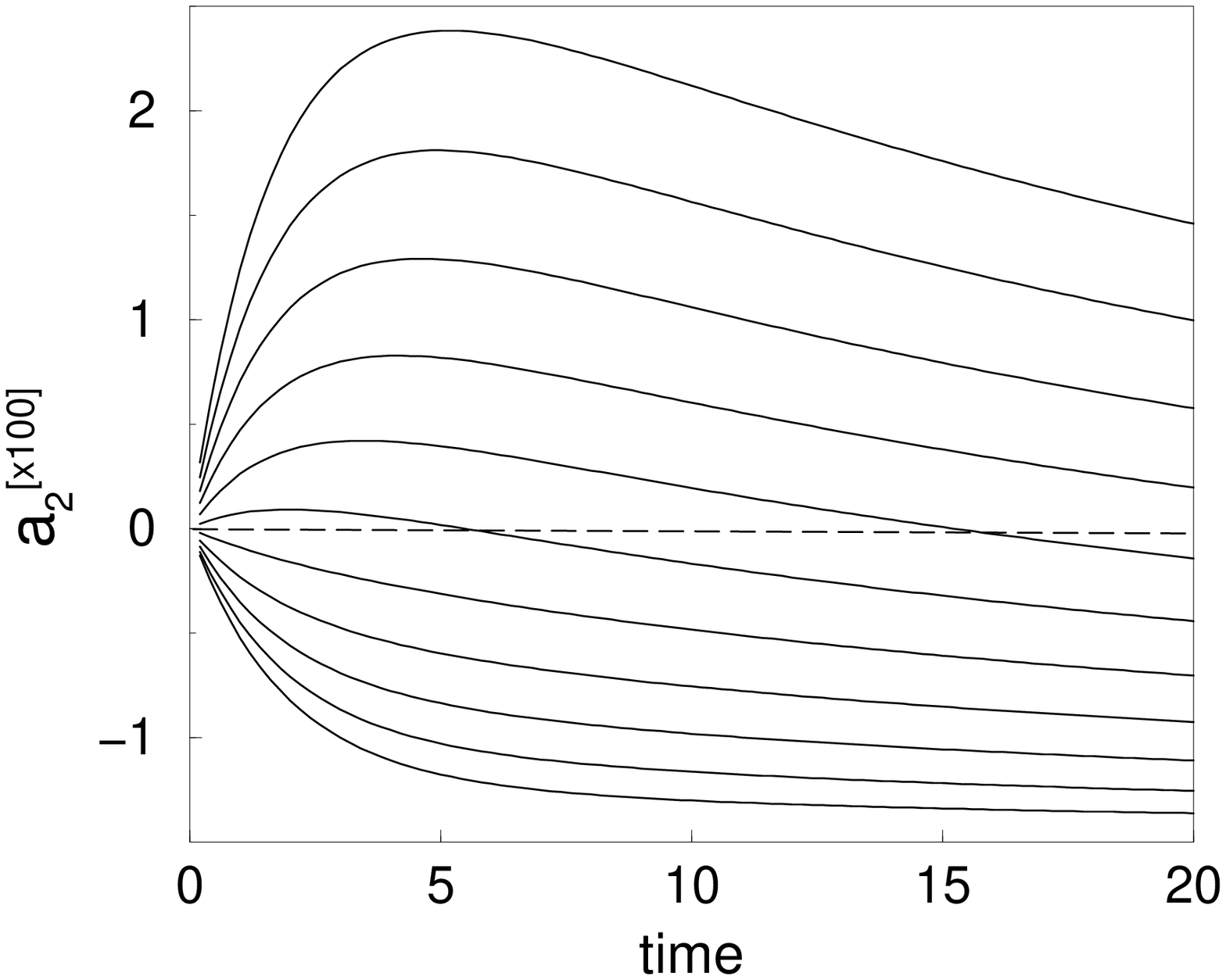,width=5.7cm}\hspace{0.3cm}\psfig{figure=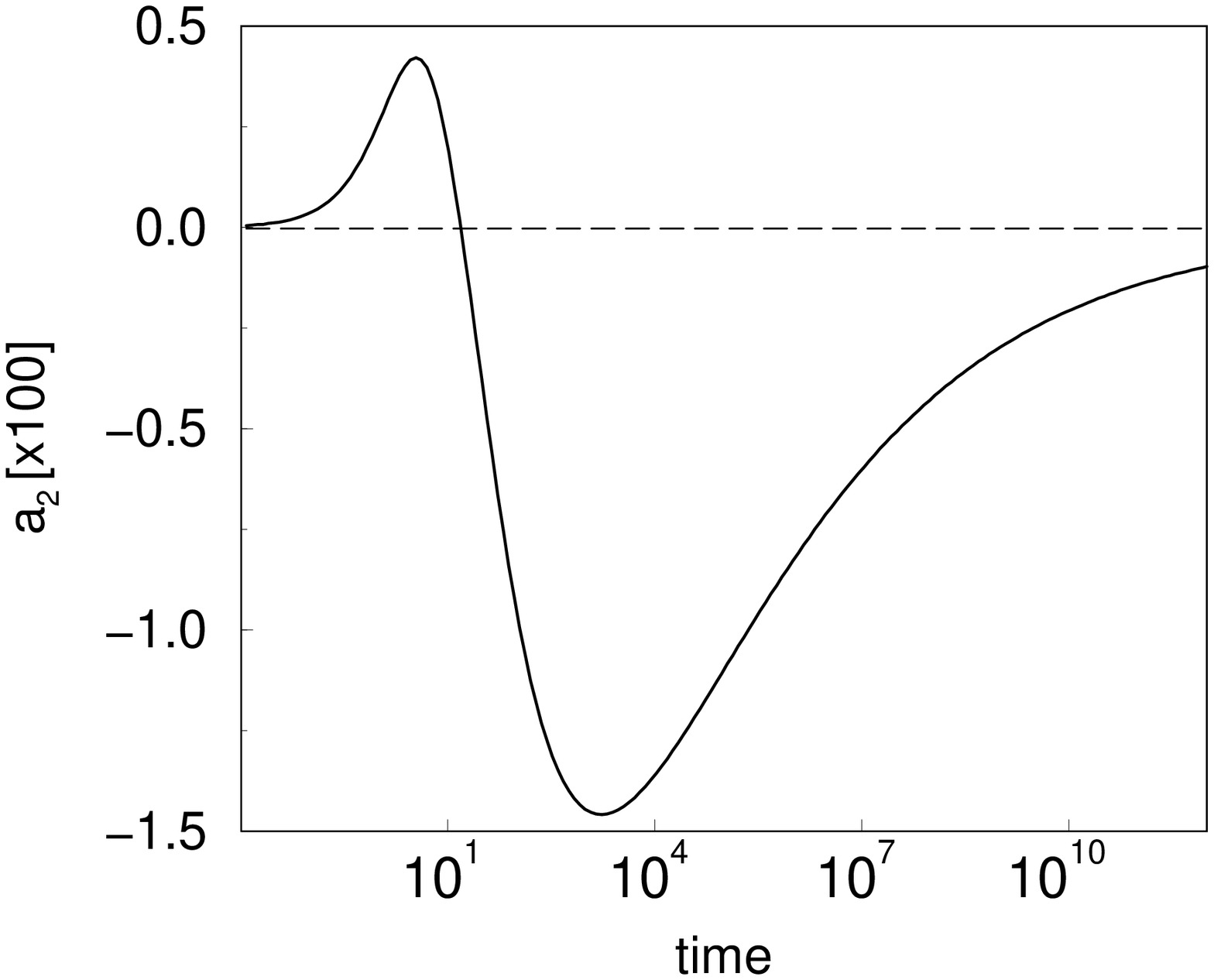,width=5.7cm}}
  \caption{The Sonine coefficient $a_2(t)$ for larger dissipation $\delta$ (numerical results). Time is given in units of mean collisional time $\tau_c(0)$. Left: $a_2(t) \times 100 $ for $\delta=0.1, 0.11, 0.12, \ldots, 0.20$ (bottom to top). Right: The plot of $a_2(t)\times 100$ for $\delta=0.16$ over logarithmic time shows all stages of
    evolution discussed in the text.}
  \label{fig:a2large}
\end{figure}

For larger values of $\delta$ the linear theory breaks down and we
performed only numerical study of the equations. The results are given
in Fig.\ref{fig:a2large}.  As compared to the case of small $\delta$, an additional
intermediate regime in the time-evolution of the velocity distribution
is observed. The first ``fast'' stage of evolution takes place, as
before, on the time scale of few collisions per particle, where
maximal deviation from the Maxwellian distribution is achieved
(Fig.\ref{fig:a2large}). For $\delta \geq 0.15$ these maximal values of $a_2$ are
positive. Then, on the second stage (intermediate regime), which
continues $10-100$ collisions, $a_2$ changes its sign and reaches a
maximal negative deviation. Finally, on the third, slow stage,
$a_2(t)$ relaxes to zero on the slow time-scale $\sim \tau_0$, just as
for small $\delta$. In Fig.\ref{fig:a2large} (left) we show the first stage of the time
evolution of $a_2(t)$ for systems with large $\delta$. At a certain
value of the dissipative parameter $\delta$ the behavior changes
qualitatively, i.e. the system then reveals another time scale as
discussed above.

\section{The age of a granular gas}
\label{sec:age}

Assume that the distribution function at time of initialization of a
granular gas is known to be Maxwellian, e.g. from the nature of the
physical process which gave rise to the granular gas. Then the
calculation in the previous section describes the evolution of the
velocity distribution function quantified by the first non-trivial
Sonine coefficient $a_2$. We have seen that this quantity evolves
characteristically in time. Assume we know an experimental method to
measure the velocity distribution over a certain time interval $(t_1,
t_2)$ and, therefore, to trace the time evolution of $a_2$ in this
interval. The time dependence of $a_2$ is known theoretically
(at least numerically) as the solution of the set of differential equations
(\ref{genseteq1},\ref{genseteq2}), which depends on the parameters  
$\tau_0$ and $\delta$, i.e. $a_2=a_2(t,\tau_0, \delta)$ 
($\tau_0$ and $\delta$, in their turn, depend on the  material
parameters $\rho$ and $A$).  This suggests a
method to {\em measure} the age of a granular gas, i.e., to determine
the time of its initialization.

\begin{figure}[htbp]
  \centerline{\psfig{figure=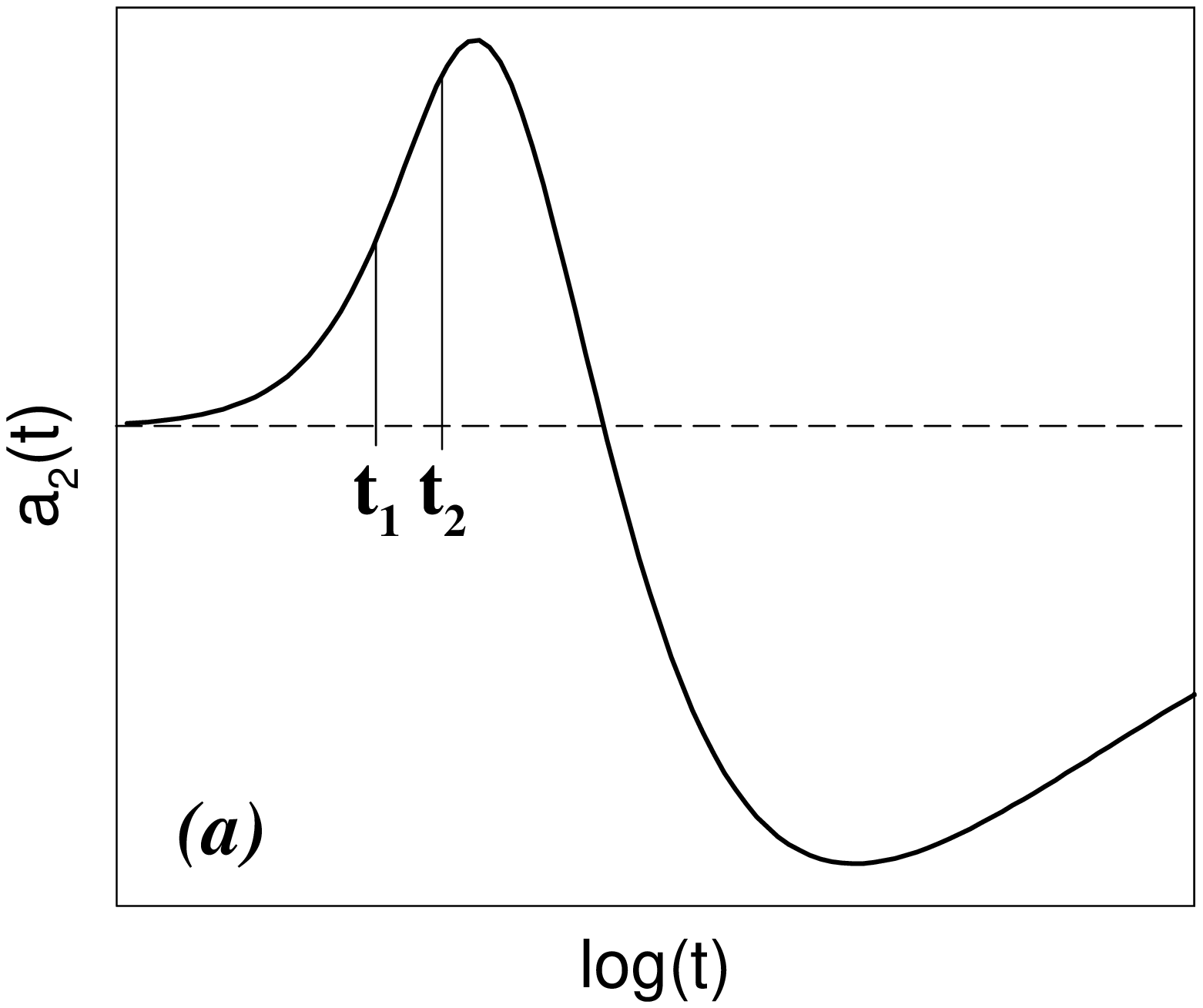,width=5.5cm}\hspace*{0.3cm}\psfig{figure=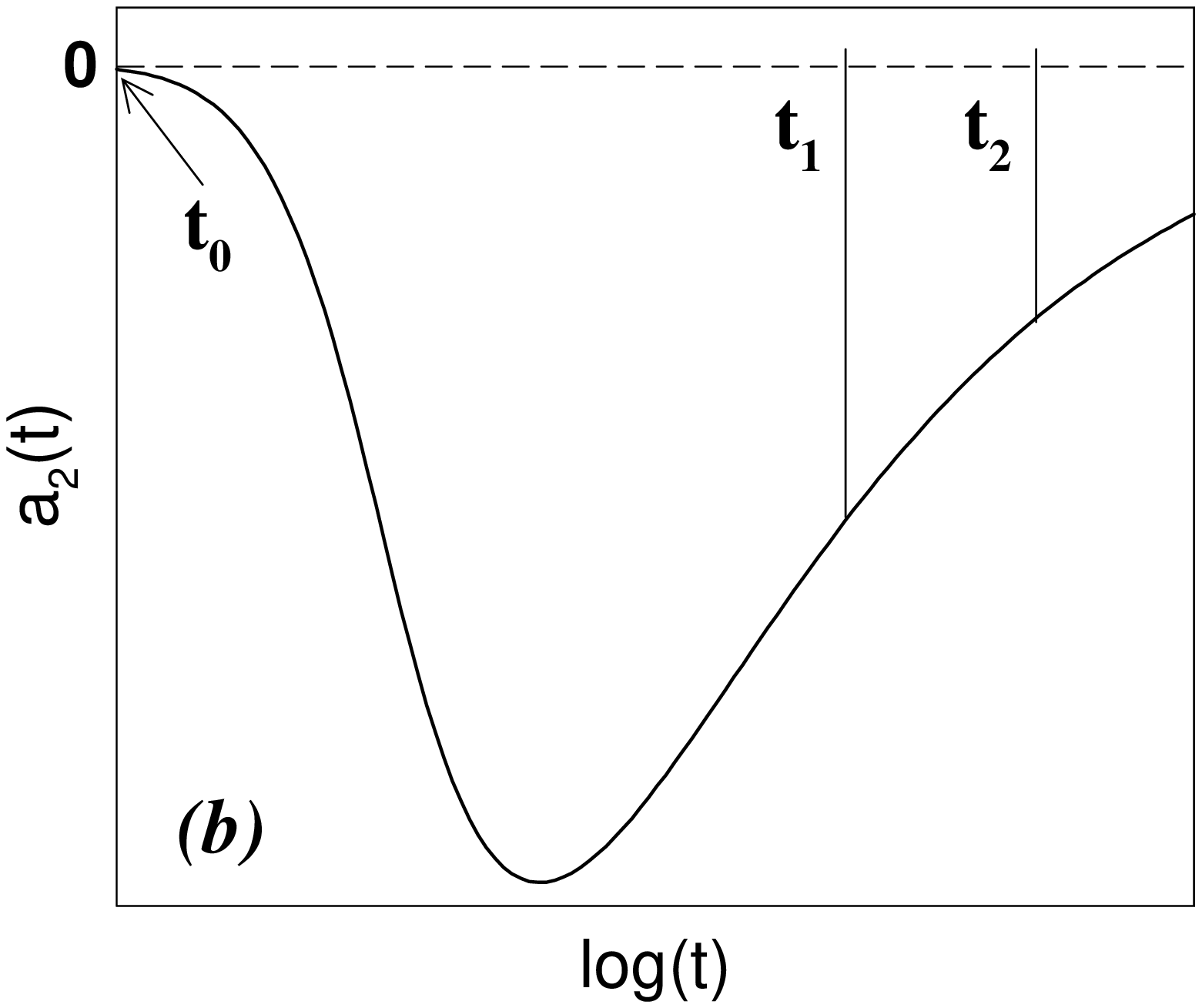,width=5.5cm}}\vspace*{0.3cm}
  \centerline{\psfig{figure=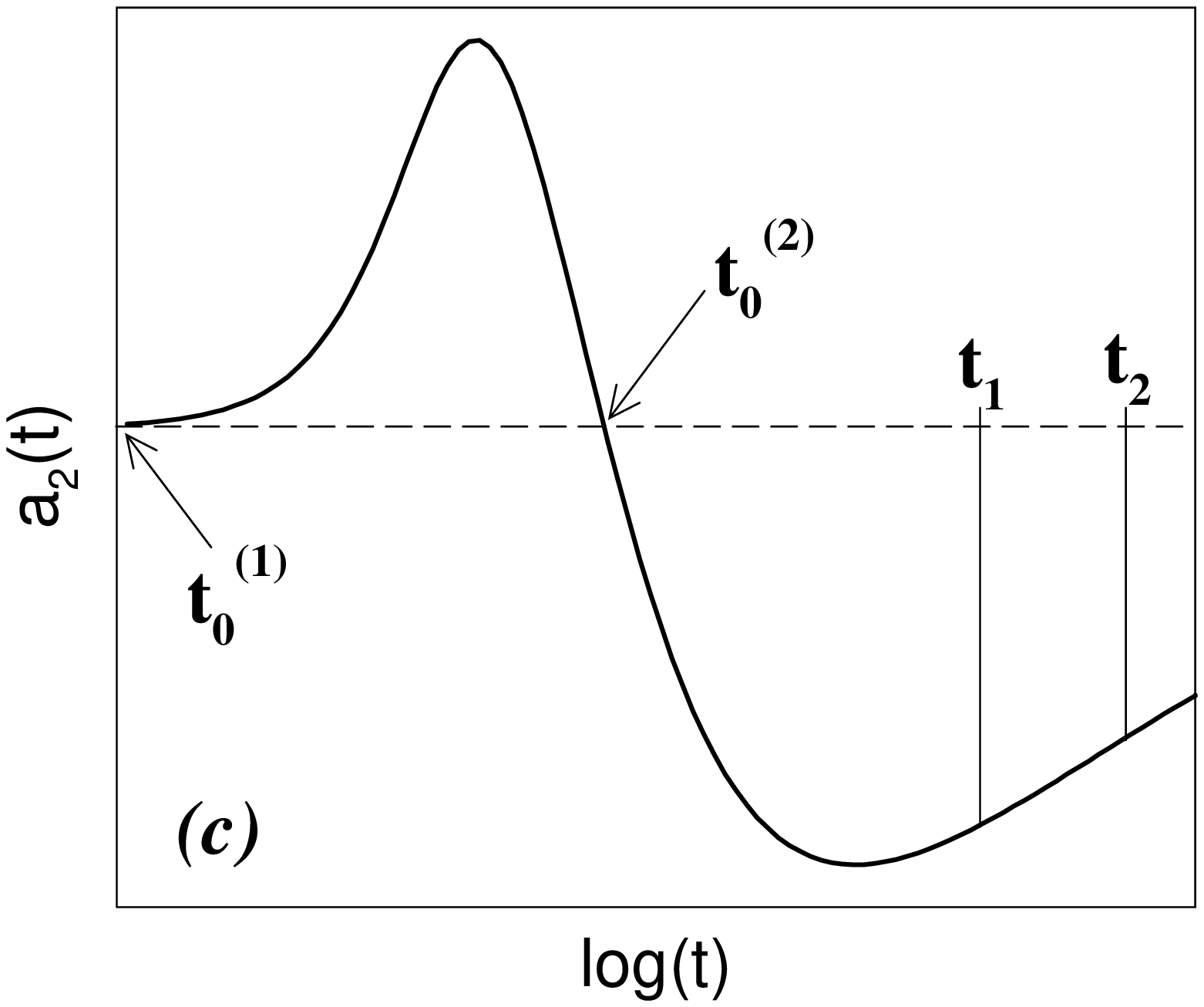,width=5.5cm}}
  \caption{Illustration of the method to compute the age of the granular gas: If the velocity distribution and, hence, $a_2(t)$  can be measured in the interval $(t_1, t_2)$, the function $a_2(t)$ can be traced backwards in time due to the described theory. If $a_2(t)>0$ for $t_1<t<t_2$ (Figure (a)), the condition $a_2(t_0)=0$ yields the age of the gas. If  $a_2(t)<0$ for $t_1<t<t_2$ depending on material properties there may be one time $a_2(t_0) = 0$ (Figure (b)) or two times for which $a_2(t_0^{(1)})=a_2(t_0^{(2)})=0$ (Figure (c)). To discriminate the cases (b) and (c) one needs further consideration (see text).
}
  \label{Fig3}
\end{figure}

Let us explain this in more detail. If we know $a_2(t)$, $t_1\le t\le
t_2$ we can compute the values of $\tau_0$ and $\delta$, which
completely parameterize the dependence $a_2(t)$, with an accuracy, which depends
on the size of the time interval $t_2-t_1$. Hence, using the solution
(either numerical or analytical) of the set of equations
(\ref{genseteq1},\ref{genseteq2}) we can trace back the dependence
$a_2(t)$ for times $t<t_1$. The time $t_0$ when the curve $a_2(t)$
cuts the abscissa corresponds to the Maxwellian distribution, i.e., it
gives the initialization time and, therefore, the age of the gas. In
Fig.~\ref{fig:a2large} (right) we see that for larger dissipation
there are two such times when $a_2\left(t_0^{(1)}\right)=
a_2\left(t_0^{(2)}\right) = 0$ ($0=t_0^{(1)} < t_0^{(2)}$ ) but only the earlier one,  $t_0^{(1)}$, corresponds to
the age (see Fig.~\ref{Fig3}). Thus, we need a method to discriminate between them: If
$a_2(t)$ was positive for at least a part of the time interval $(t_1,
t_2)$, then the initialization time $t_0$ is uniquely determined (Fig.~\ref{Fig3}a). If
the value of $a_2$ was negative (Fig.~\ref{Fig3}b,c) we can trace time-backwards the
dependence $a_2(t)$ until at time $t_0$ the condition $a_2(t_0)=0$ is
fulfilled. Then additional analysis is required: The value of $\delta$
which was already computed determines the ``steady-state'' value of $a_2$ which is reached after
the quick relaxation in a time of the order of few mean-collision
times. For small dissipation $\delta$ it is negative and for large
dissipation it is positive. Thus, if the steady state value of $a_2$ is
negative, the time $t_0$ with $a_2(t_0)=0$ corresponds to the
initialization time, i.e. $t_0=t_0^{(1)}$. Otherwise 
(i.e. for positive steady-state $a_2$), $t_0$ corresponds to $t_0^{(2)} >0$, and 
one has to  trace $a_2(t)$ further
time-backward in order to find the next time $t_0$
which fulfills $a_2(t_0)=0$ to find the time of initialization.

If the age of the gas has been measured according to the described
method one can also calculate the initial temperature $T_0$, i.e. the
initial energy of the gas. From $\delta$ and $T_0$ according to the
definition of $\delta$ one can estimate the combination of the
material parameters $A\alpha^{2/5}$, and even the size of the
particles $\sigma$ using Eq.(\ref{tau0}), which relates $\tau_0$,
$T_0$, $\delta$ and $\sigma^2 g_2(\sigma)$, provided the number
density $n$ may be measured.

Assume an astrophysical catastrophic impact took place at a certain
time and produced a granular gas cloud with Maxwellian velocity
distribution. If one would be able to measure the velocity
distribution function in a much later time interval $t_1 \le t\le t_2$
following the described procedure one would be able to determine (i)
the time when the impact took place, (ii) the energy of the impacting
bodies (from the initial temperature), 
(iii) some material properties of the bodies and (iv) the
grain size of the granular gas. The described analysis does not
require the knowledge of the material properties of the particles.

\section{Conclusion}

We analyzed the time evolution of the velocity distribution function
in a granular gas of viscoelastic particles in the homogeneous cooling
state. The assumption of viscoelasticity is the simplest assumption
for the dissipative collision of particles which is in agreement with
mechanical laws. The collision of these particles is characterized by
an impact-velocity dependent restitution coefficient.

For the case of a gas of particles which interact via a constant
restitution coefficient its evolution is completely determined by the
time dependence of the temperature. The velocity distribution function
has a simple scaling form, i.e., it depends only on
the reduced velocity of the particles, $\vec{c}=\vec{v}/v_0(t)$, which
is just the velocity measured in units of the characteristic velocity
$v_0(t)$, related to the current temperature. The scaling form, thus,
persists with time.

Contrary, the velocity distribution function of a gas of viscoelastic
particles does not have a simple scaling form. The deviation of the
velocity distribution from the Maxwellian which is for the case
$\epsilon=\mbox{const.}$ a function of $\epsilon$ only, i.e.
time-independent, depends for a gas of viscoelastic particles
explicitely on time, i.e. the velocity distribution function undergoes
a time evolution. We quantify 
the deviation from the Maxwellian
distribution by means of the first non-vanishing term of  the Sonine polynomials 
expansion, characterized by the coefficient $a_2$.  We assume that inelasticity of the particles is small and, hence, higher order terms may be neglected. Contrary to the case of the constant restitution coefficient, where
$a_2={\rm const.}$, for a gas of viscoelastic particles $a_2$ reveals
a rather complex time behavior with different regimes of evolution.

The time dependence of the distribution function,  quantified by $a_2(t)$, 
exhibits different stages of evolution, which allows
to assign a Granular Gas the term ``aging''. The explicite time
dependence implies that the process has a definite starting point of
initialization, i.e. earlier times do not correspond to a physically
meaningful state of the system. At first glance it might appear
cumbersome that a granular gas, i.e. a cooling gas of dissipatively
colliding particles, has a well defined initialization time. Indeed,
if the temperature was the only characteristics of the system there
would be no reason to mention some specific initial temperature, since
the gas may exist at any temperature. In this case we would not be able 
to conclude whether the present state of a gas is the starting point of its
evolution or an intermediate one, so that its history started long ago at 
 much higher temperature. This is the case if  the particles interact via 
a constant restitution coefficient 
$\epsilon=\mbox{const.}$  For gases of viscoelastic particles, however,
the velocity distribution evolves in a way which allows to determine univocally the
time-lag from the starting point and, hence, the age of the granular
gas.

{F}rom the knowledge of the velocity distribution in a certain time
interval $t_1\le t\le t_2$ we can determine not only the age of the
gas but also its initial temperature and certain material properties
of the particles. This result may be useful to determine the time, the
energy and other system properties of catastrophic impacts in
astrophysical systems.

\end{document}